\documentclass[iop]{emulateapj}

\begin{document}

\title{The Second Arecibo Search for 5 GHz Radio Flares from Ultracool Dwarfs}

\author{Matthew Route\altaffilmark{1,2,3} \& Alexander Wolszczan\altaffilmark{1,2}}

\altaffiltext{1}{Department of Astronomy and Astrophysics, the Pennsylvania State University, 525 Davey Laboratory, University Park, PA 16802, alex@astro.psu.edu}

\altaffiltext{2}{Center for Exoplanets and Habitable Worlds, the Pennsylvania State University, 525 Davey Laboratory, University Park, PA 16802}

\altaffiltext{3}{Current Address: Research Computing, Information Technology at Purdue, Purdue University, 155 S. Grant St., West Lafayette, IN 47907, mroute@purdue.edu}

\slugcomment{Accepted for publication in ApJ: 2016 August 8}

\begin{abstract}
We describe our second installment of the 4.75 GHz survey of ultracool dwarfs (UCDs) conducted with the Arecibo radio telescope, which has observed 27 such objects and resulted in the detection of sporadic flaring from the T6 dwarf, WISEPC J112254.73+255021.5. We also present follow up observations of the first radio-emitting T dwarf, 2MASS J10475385+2124234, a tentatively identified radio emitting L1 dwarf, 2MASS J1439284+192915, and the known radio-flaring source, 2MASS J13142039+132011 AB.  Our new data indicate that 2MASS J1439284+192915 is not a radio flaring source.  The overall detection rate of our unbiased survey for radio-flaring UCDs is $\sim$5\% for new sources, with a detection rate for each spectral class of $\sim$5-10\%. Evidently, radio luminosity of the UCDs does not appear to monotonically decline with spectral type from M7 dwarfs to giant planets, in contradiction to theories of the magnetic field generation and internal structure of these objects.  Along with other, recently published results, our data exemplify the unique value of using radio surveys to reveal and study properties of substellar magnetic activity.
\end{abstract}
 
\section{Introduction}

Ultracool dwarfs (spectral types $>$M7, including L, T, and Y) are examined as a group on account of their common interior structure and displayed magnetic phenomena.  Low-mass stellar evolutionary models demonstrate that beyond spectral type $\sim$M3 (0.35 M$_{\odot}$), stellar interiors should become fully convective \citep{cha97,cha00}.  However, in the absence of a radiative zone, the tachocline, the shearing interface between the radiative core and the convective layer thought to be the key component in generating a large-scale magnetic field, ceases to exist as well, making the generation of strong magnetic fields in these low-mass, generally rapidly rotating objects problematic.  Several dynamo mechanisms have been proposed to account for this difficulty, including turbulent, $\alpha^{2}$, and $\alpha^{2}\Omega$ models (\citet{cha06} and references therein).  As UCDs bridge stars and planets, the functioning, properties, and manifestations of their magnetic dynamos may overlap with both stellar dynamos and geodynamos \citep{bro08,chr09}.  Thus, studies of UCDs are important to the efforts to improve both sets of magnetohydrodynamic models.

The dynamo-generated magnetic activity found among UCDs is manifested through various indicators, including X-ray, near-infrared/optical, and radio emission.  All three are useful for diagnosing magnetic strength and topology on the Sun and other stars, while radio emission has shown its value in permitting the study of the magnetospheres of the Solar System giant planets \citep{zar98}.  However, such emission has yet to be found among extrasolar giant planets (e.g. \citet{laz07}).  Various near-infrared/optical spectral lines, such as Ca II H and K, H$\alpha$, and the coronal green line (Fe XIV 5303 $\AA$, \citet{mci14}) have been successfully used to trace magnetic structures in the solar chromosphere and corona, while FeH has been used in low mass stars to measure their magnetic flux \citep{rei07}.  However, UCDs generally rotate rapidly, causing such spectral lines to broaden to the degree that the signature of Zeeman splitting is obscured.  X-ray emission appears to be suppressed beyond spectral type M7 (e.g. \citet{mcl12}), perhaps due to the decoupling of magnetic fields from neutral atmospheres \citep{moh02}, although more recent calculations indicate that this should not pose difficulties in the generation of magnetic activity for spectral types before $\sim$L4 \citep{rod15}. Alternatively, centrifugal effects may cause reduced H$\alpha$ and X-ray activity, due to the concentration of magnetic fields toward the poles, thus reducing the field filling factor, or causing the stripping of the corona (\citet{ber08b} and references therein).  This leaves H$\alpha$ and radio activity, which appear to be correlated \citep{kao16}, as the indicators available to use to study the magnetic fields of UCDs.

Radio emission detected from the first radio-loud UCD, the M9 dwarf LP 944-20, consisted of both flaring and non-flaring components. All of it was interpreted as gyrosynchrotron emission due to its broadband character, non-thermal brightness temperature (T$_b\sim 4\times 10^{9}$ K during flares), and low circular polarization fraction ($\sim$30\%) \citep{ber01}.  The magnetic field strengths corresponding to this mechanism are B$\sim$10 G for this source, although on the Sun such a mechanism causes solar flares from active regions with B$\sim$500 G \citep{whi11}.  On the other hand, \citet{hal06,hal07} detected periodic radio emission from the M9 dwarf TVLM 513-46546 and reasoned that the highly directed emission must come from emitting regions smaller than the surface of the disk, thereby resulting in T${_b}> 2.4\times 10^{11}$ K.  The improved temporal resolution of the data demonstrated that the periodic radio signal consisted of 100\% circularly polarized bursts that, together with the coherence of the emission, pointed to the electron cyclotron maser instability (ECMI; \citet{tre06}) as the most likely generation mechanism. ECMI-induced radio-emitting sources have much stronger $\sim$kG magnetic fields, as determined by the cutoff frequency in the radio emission, the lower bound of which corresponds to the local cyclotron frequency (Hz), $\nu_{c} = 2.8 \times 10^{6} B$ (Gauss).  This mechanism is the leading theory to explain the generation of the auroral kilometric radiation (AKR) at the Earth, as well as similar radio emission for Jupiter and Saturn \citep{zar98}. The emission from TVLM 513-46546 also consisted of a quiescent, non-flaring component, that \citet{hal08} attributed to either gyrosynchrotron or depolarized ECMI emission.  However, as \citet{wil14} suggested for 2MASS J07464256+2000321 AB (hereafter, J0746+20 AB), it is also possible that for UCDs with both quiescent and flaring radio emission, the former is induced by gyrosynchrotron emission, while the latter results from ECMI.

Inspired by the detection of both quiescent and flaring radio emission from LP 944-20, a number of surveys have been conducted to search for similar objects. \citet{ber02} conducted a survey of 12 late M and L dwarfs using the Karl G. Jansky Very Large Array (VLA), detecting three new sources. \citet{bur05} used the Australia Telescope Compact Array (ACTA) to examine seven late M and L dwarfs, detecting flaring and quiescent emission from two sources.  A much larger VLA survey of 90 stars ranging in spectral type from M5 to T8 detected three new sources \citep{ber06}, while a later survey of 104 similar targets detected three new sources \citep{mcl12}.  A small VLA survey of late M dwarfs detected quiescent emission from a tight M8 binary \citep{pha07}.  Another VLA mini-survey of eight UCDs of spectral classes M8.5 to T6 detected the most radio-luminous brown dwarf to date, the binary L0.5+L1.5 system J0746+20 AB \citep{ant08}, although a later survey of 32 sources encompassing spectral types M7 to T8 detected no new sources \citep{ant13}.  The accumulated detection statistics up to that time suggested an overall radio detection efficiency of $\sim$9\% for M7 to L3.5 UCDs. The first Arecibo survey of UCDs \citep{rou13}, examined 34 objects of spectral types M9 to T9, resulting in the expansion of known radio-emitting UCDs to include T dwarfs, with the detection of the radio flaring T6.5 source, 2MASSI J10475385+2124234 (J1047+21; \citet{rou12}).  This latest survey was only sensitive to rapid, flaring radio emission, and computed a $\sim$7\% detection rate for flaring emission.  Another recent mini-survey of 15 late M and L dwarfs using ATCA resulted in the detection of one new quiescent source \citep{lyn16}.  More recently, a small, biased, upgraded VLA survey of six L and T targets with known near-infrared/optical variability had a success rate of 80\% \citep{kao16}.

Our latest radio survey searching for flaring L and T dwarfs has been motivated by our success in detecting the coolest radio-flaring UCD to date, J1047+21 \citep{rou12}, and our desire to find more such sources to better illuminate the radio-emitting characteristics of the brown dwarf population, especially toward the less-well studied later spectral types. In Section 2, we describe and assess the capabilities of the Arecibo radio telescope, and discuss the composition of our latest survey target list. The results from our survey, with a focus on key objects of interest are presented in Section 3, while the statistical significance of these results and the information they suggest about trends in UCD magnetism are discussed in Section 4.  Section 5 summarizes the salient points of our survey, and provides suggestions for future developments in the field.

\section{Target Selection and Observations}

The observations described here mark the second installment in our UCD survey conducted with the Arecibo radio telecope, as initially described in \citep{rou13}. So far, the survey has led to two new detections of radio emitting brown dwarfs, including the T6.5 UCD 2MASS J10475385+212423 (J1047+21; \citep{rou12}), which is the coolest known radio emitting UCD, and the T6 UCD WISEPC J112254.73+255021.5 {J1122+25; \citep{rou16}), which is likely to be the most rapidly rotating UCD discovered.  Previous radio surveys have exploited differing benefits of the instruments that they utilized: the surveys presented in \citep{ant08,ant13,ber02,ber06,mcl12,pha07} have leveraged the VLA's ability to detect both quiescent and flaring radio emission, although integration times are $\sim$10 s or larger.  The results from our surveys leverage Arecibo's complementary abilities to measure rapid, $\sim$0.1 s, source variability at a high signal-to-noise ratio, due to the increased sensitivity that results from the large collecting area of its 305 m dish.  However, while Arecibo observations accurately record rapid changes in flux density, the instrument is insensitive to the detection of quiescent emission due to its usage of a calibration procedure that relies on a locally generated calibration signal as opposed to the quasar calibration sources used at VLA (e.g. \citep{wilb15}).  While the local calibrator does not provide the absolute flux density, it does have the advantage that only 20 s are used for calibration purposes after a 600 s scan, while VLA observations require anywhere from $\sim$15\% \citep{hal06} to over one-third of the observing time to be spent in calibration \citep{wil14}.  This makes it less likely that short radio bursts will be missed or incompletely characterized with the Arecibo radio telescope.

Twenty-seven ultracool dwarfs were observed during the course of this survey, 20 of which were previously unobserved with Arecibo, and 19 of which have not been observed previously by any radio telescope (see Table 1).  We chose our targets based on both their intrinsic physical properties and in accordance with observing constraints.  Our targets were all located $<$25 pc and had spectral types ranging from M7 to T9, with the majority of targets of class T1 or later.  Due to the fixed-dish nature of the Arecibo radio telescope, all objects had declinations of 0$^\circ$ to +38$^\circ$ and due to scheduling constraints, right ascensions between 7 to 14 hr.  All sources were observed for $\sim$2 hr, mostly in a contiguous block, which represents the time required for a target to transit the fixed dish.  The only exception was J1122+25 which was repeatedly observed in $\sim$2 hr blocks in accordance with our efforts to verify its flaring nature and obtain a rotational period for the radio emission, as described in \citet{rou16}.

Eight targets that had been previously observed by various radio telescopes were re-observed here for several reasons.  The sources 2MASS J07003664+3157266 A, 2MASS J07271824+1710012, SDSS J082519.45+211550.3, 2MASS J09373487+2931409, and ULAS J133553.45+113005.2 were observed in \citet{rou13}, but were re-observed on account of the possible variability in radio activity, as demonstrated by \citet{wilb15}.  Follow up observations were conducted of J1047+21, for which we detected three radio bursts \citep{rou12} but subsequent efforts permitted both the detection of quiescent radio emission \citep{wil13} and the determination of its rotational period \citep{wilb15}, as well as 2MASSW J1439284+192915 (J1439+19) which we previously noted was a potential radio-flaring source \citep{rou13}.  Finally, although not previously detected with the Arecibo radio telescope, 2MASS J13142039+1320011 AB (J1314+13 AB) was also examined to study its flare profiles at high temporal resolution \citep{wil15}.

\footnotetext[1]{ ``C-Band,'' available at http://www.naic.edu/$\sim$astro/RXstatus/Cband/Cband.shtml}

These observations were conducted between 2013 March 5 and 2014 January 5 using the C-band receiver and Mock spectrometers.  The antenna gain and system temperatures were 6 to 9 K Jy$^{-1}$ and $\sim$30 K, respectively\footnotemark.  The C-band receiver has a center frequency of 4.75 GHz and a half-power beam width of approximately 1 arcmin in both azimuth and zenith angle.  The receiver passes dual-linear polarization signals to an array of seven field-programmable gate array- (FPGA) equipped Fast Fourier Transform (FFT) Mock spectrometers, each of which has a 172 MHz bandpass divided into 8192 channels, arranged to yield an $\sim$1 GHz simultaneous bandpass \citep{sal09}.  The Mock spectrometers were sampled at 0.1 s intervals.  Our natively developed software described in \citet{rouphd}, computed the Stokes parameters, performed flux calibration, bandpass correction, radio frequency interference (RFI) removal, and resampled the data to $\sim$80 kHz spectral and 0.9 s temporal resolution for further study.  The software performance and our analysis approach were validated via observations of flares in TVLM 513-46546 \citep{ber02,ber08a,hal06,hal07,ost06} and 2MASSW J0746425+200032 AB \citep{ant08,ber09}, as previously described in \citet{rou13,wol14}.

Due to the nature of our instrumental set up, this survey was sensitive to rapid flares with rise times $\sim$mins or less, that were at least modestly ($>$ 10\%) circularly polarized.  In addition, while the theoretical 1$\sigma$ sensitivity is $\sim$0.15 mJy, in practice this rises to 0.4- 1.6 mJy due to the presence of RFI across all sub-bands, with the lowest (centered at 4.35 GHz) and highest (centered at 5.35 GHz) sub-bands being especially noisy and growing increasingly problematic with time. 

\subsection{Noteworthy Re-observed Objects}
In particular, three radio-loud UCDs require further description: 2MASSI J10475385+2124234, 2MASS J13142039+1320011 AB, and 2MASSW J1439284+192915.  J1047+21 was discovered during a search for objects cooler than Gl 229B in the 2MASS Spring 1999 Data Release, that had no optical or minor planet counterparts, but J$<$16, J-H$<$0.3, and H-K$_{S}<$0.3 \citep{bur99}.  \citet{bur02} categorized the brown dwarf as a T6.5 dwarf based on H$_{2}$O and CH$_{4}$ spectral indices derived from NIRC near-infrared spectroscopic data.  US Naval Observatory astrometry permitted \citet{vrb04} to derive a 10.6 pc distance to J1047+21, using trigonometric parallax, as well as T$_{eff}\sim$870 K. \citet{bur03} measured marginally significant H$\alpha$ emission using LRIS red optical spectra.

Initial VLA radio observations at a center frequency of 8.46 GHz established a 45 $\mu$Jy upper limit to its quiescent radio flux density at 8.46 GHz \citep{ber06}. Later Arecibo observations by \citet{rou12} at 4.86 GHz demonstrated that the brown dwarf was a sporadic radio emitter through the detection of three flares with circular polarization fractions ranging from 18 - 89\%, and $T_{B}>$10$^{11}$ K.  These emission properties are widely thought to be the result of ECMI \citep{ber09,hal06,hal08,rou12,rou16,wil15}, which allowed for the determination of a lower bound to the magnetic field strength of $B>$1.7 kG. The novel usage of dynamic spectra as applied to brown dwarf radio emission resulted in the detection of drifting radio emission in frequency space, which, for solar-like plasma emissions, indicated a magnetic field scale size of $a\approx$ 0.3 R$_{J}$ - 1 R$_{J}$.  This flaring emission was quickly followed up by the detection of persistent radio emission using VLA at 5.8 GHz \citep{wil13}. This group detected no variability in the 16.5 $\mu$Jy emission, established an upper bound to the circular polarization ($<$80\%), and computed a low brightness temperature ($T_{B}\sim$10$^{8}$ K), indicating its gyrosynchrotron nature. Later VLA observations of J1047+21 at a center frequency of 6 GHz measured a flaring periodicity of $\sim$1.77 hr from several highly circularly polarized flares ($\sim$50-100\%), with significant variability in their amplitudes \citep{wilb15}.  During these observations, a single pulse at 10 GHz was observed as well, suggesting that $B\gtrsim$3.6 kG in certain emitting regions.  \citet{kao16} verified these results with enhanced VLA observations that detected both quiescent and flaring radio components at $\sim$6 GHz.  By applying the substellar evolutionary models of \citet{bar03}, they derived a system age $>$2.5 Gyr, and a mass $>$0.026 M$_{\odot}$.

The M7 dwarf 2MASS J13142039+1320011 AB ($=$ NLTT 33370 AB; J1314+13 AB) is a young, active binary system initially detected in the New Luyten Two-Tenths catalog as a high-proper motion object \citep{luy79}. Early spectroscopic measurements revealed that the system strongly emits H$\alpha$, with L$_{H\alpha}$/L$_{bol}\approx$ -3.2 \citep{lep09}. \citet{law06} analyzed {\it i} and {\it z} band observations with the Lucky Imaging technique, exposing the binary nature of the source, the two components of which are separated by 0.13 arcsec, or $\sim$2.1 AU.  Multi-epoch Very Long Baseline Array (VLBA) data revealed that the secondary is the source of the radio emission and provided a distance measurement of 17.249 $\pm$ 0.013 pc \citep{for16}.  These VLBA observations, combined with Keck adaptive optics near-infrared camera (NIRC2) astrometry and near-infrared optical spectroscopy, permitted \citet{dup16} to measure the component masses as 92.8 $\pm$ 0.6 M$_{J}$ and 91.7 $\pm$ 1.0 M$_{J}$, with T$_{eff}$ of 2950 $\pm$ 5 K and 2770 $\pm$ 100 K, respectively.

\citet{mcl11} first reported its detection as a $\sim$1 mJy variable radio source at 1 - 22 GHz frequencies in VLA observations.  The 20- 30\% amplitude, sinusoidal variation at 4.86 and 8.46 GHz indicated a periodic radio source with period $P$ = 3.89 hr. The low polarization of the radio emission ($\pm$24\%), coupled with a nearly 20 GHz emission bandwidth are indicative of a gyrosynchrotron emission mechanism \citep{gud02}.  Subsequent VLA observations \citep{wil15} revealed the presence of strong ($<$ 8 mJy), several minute-duration, 100\% circularly polarized flares superimposed upon the previously detected quiescent emission.  Such high brightness temperature, highly polarized flares indicated the ECMI process, and therefore describe an emitting region magnetic field strength $B$ $\sim$ 2.1 kG.

2MASSW J1439284+192915 (J1439+19) was discovered during a search of the 2-Micron All-Sky Survey (2MASS) catalog for cool, reddish sources with K$_{S}$ $\leq$ 14.50 and no optical counterpart \citep{kir99}.  They used J1439+19 to define the L1 spectral subclass on account of the nearly equal strength of TiO, CrH, and FeH features, the strengthening of Rb I and Cs I absorption, and the slight sloping of the 7800 - 8000 $\AA$ portion of the spectrum.  However, no H$\alpha$ emission was detected.  Measurements of the L1 dwarf's radial velocity using NIRSPEC yielded a radial velocity of $v$sin$i$, of 11.1 km s$^{-1}$ \citep{bla10}.  Observations conducted by \citet{dah02} during their US Naval Observatory CCD faint star parallax program determined a distance to J1439+19 of 14.4 pc on account of its trigonometric parallax, which would result in T$_{eff}\sim$ 2270 K.  During a Hubble Space Telescope Snapshot project that searched for L dwarf binaries, \citet{rei08} failed to find any companion to J1439+19 within 0.3 arcsec (4.32 AU) for mass ratios q $>$ 0.2. While \citet{mcl11} described J1439+19 as a radio-quiet source, with a 78 $\mu$Jy radio flux density upper limit, \citet{rou13} tentatively reported the detection of a $\sim$1 mJy, 90\% circularly polarized, $\sim$5 GHz flare during the course of their first Arecibo UCD radio survey.  These emission characteristics suggested an ECMI mechanism, coming from an emission region with $B$ $>$ 1.5 kG.

\section{Results}

This latest survey of UCDs has yielded the detection of radio flares from three targets, including the detection of a new source, the T6 dwarf J1122+25, as well as the observation of flares from two previously detected UCDs, the M7 binary J1314+13 AB, and the T6.5 dwarf J1047+21. J1122+25 is only the fourth detected radio emitting T dwarf, with an inferred effective temperature of $\sim$1060 K \citep{vrb04}.  With a cutoff frequency of $>$5.2 GHz, it has a flaring radio luminosity $\nu$L$_{\nu} > 5.1 \times 10^{24}$ ergs s$^{-1}$, making it the most radio luminous T dwarf at this time, being nearly three times more radio luminous than J1047+21. It is also the most energetic radio-emitting, isolated brown dwarf detected thus far.  Our capture of several burst events during our observation campaign has led to a calculation of its rotational period of 0.288 hr, with the second and third subharmonics of this period also being acceptable solutions to the temporal data, any of which make this the fastest rotating brown dwarf measured yet \citep{rou16}.  The large radio luminosity could very well be a result of the object's ultra-rapid rotation. 

Our observing program also detected a $\sim$1 mJy (5.7$\sigma$ significance) $\sim$57\% left circularly polarized burst from J1314+13 AB, as depicted in Figure 1. This flare is roughly in agreement with the characteristics reported by \citep{wil15} that indicated an ECMI origin to the flares. If this emission mechanism caused the flare, the presence of RFI at higher frequencies only allows us to constrain the computed magnetic field strength to be B$\sim$1.7 kG.  The dynamic spectra, although polluted with RFI, indicate a frequency drift rate of $d\nu/dt\sim$30--700 MHz s$^{-1}$, with a smaller drift rate at lower frequencies.  Thus, given the observations of quiescent radio emission by \citet{mcl11}, this object supports the notion that at least some UCDs display radio emission consistent with both gyrosynchrotron and ECM emission.

Our follow-up observations of J1047+21 resulted in the detection of only a single $\sim$1.9 mJy flare, of $\sim$30 s in duration, peaking in intensity at MJD 56418.987053.  This flare is 33\% right-circularly polarized, with a cutoff frequency of $\sim$5.0 GHz, yielding a maximum magnetic field strength of $\sim$1.8 kG (Figure 2).  For an emission region of $\sim$1 R$_{J}$ in size, we determine a brightness temperature of T$_b\geq 2\times 10^{10}$ K, placing it in the regime of electron cyclotron maser emission. Although this polarization fraction is smaller than many other detected ECM flares from similar UCDs \citep{ant08,hal06,hal08,wil15}, the polarization is at the upper end of what would be anticipated for gyrosynchrotron emission, while having a much larger brightness temperature.  This flare may therefore be a result of ECMI, but with the emission depolarized as it travels through intervening plasma, a hypothesis advanced in \citet{hal08} to explain the low degree of circular polarization of the interpulse quiescent radio emission observed from TVLM 513-46546.

As part of this observing program, we also re-observed J1439+19, which appeared to emit radio bursts during three of five observing sessions that were greater than 30 mins in length, spanning 2013 May 3 to 2015 May 14.  The radio bursts each had flux densities of 1.75 - 1.80 mJy, with significance of 3.5 - 4.1$\sigma$.  All bursts were observed at nearly the same time of the day, with right circular polarization fraction varying from $\sim$30\% to $\sim$100\%. Moreover, based on the best-fit period to these three events, we compute a period within $\sim$0.6 s of the length of the sidereal day, suggesting that the emission is not stellar in nature, but rather points to a local phenomenon.  We therefore conclude that the flaring behavior observed to be associated with J1439+19 is most likely terrestrial RFI.  This conclusion, coupled with the fact that previous work by \citet{kir99} failed to detect H$\alpha$ emission, appears to demonstrate that J1439+19 is magnetically inactive, and supports the linking of the two types of emission as found in \citet{kao16}.

Of the 19 newly observed brown dwarfs, only one new source, J1122+25 was detected, yielding a detection rate of $\sim$5\% for new sources, and $\sim$4\% for the 24 sources which includes both new and revisited sources without a prior detection.  Table 2 presents the maximum radio burst flux density (or upper limits) for detected (non-detected) objects from both the 20 newly observed UCDs and the revisited objects from previous surveys.  These upper limits are derived from the 3$\sigma$ standard deviation of the frequency-integrated time series from the cleanest sub-band, centered at 4.47 GHz, when the timing resolution has been smoothed to 0.9 s.  Of the sources 2MASS J07003664+3157266 A, 2MASS J07271824+1710012, SDSS J082519.45+211550.3, 2MASS J09373487+2931409, and ULAS J133553.45+113005.2 that were re-observed from our previous program, no flares were detected from any of them. In fact, we note that our upper detection limits for these sources are systematically less sensitive than in our earlier survey conducted from 2010 to 2013 \citep{rou13}. This reflects the increasing noisiness of the radio environment surrounding Arecibo and the necessity of restricting external interference in scientific observing bands. Although in Table 2 we report the results from this program, in Figures 3 and 4 we plot the most sensitive results for these objects (i.e. from our previous survey).

\section{Discussion}

\subsection{UCD Radio Flaring Observing Statistics}

This survey attempted to improve on the detection performance of our previous survey \citep{rou13}, which had a reported detection efficiency of $\sim$7\% (but an actual efficiency of $\sim$3\% given our improved knowledge of J1439+19) for objects with rapidly varying, polarized radio bursts, principally through the reduction in distances to sources from $<$40 pc to $<$25 pc.  However, our detection efficiency of $\sim$5\% for our not-previously detected targets does not represent an improvement over previous surveys, as summarized by \citet{rou13} and \citet{lyn16}.  Both publications reveal that the detection rate of unbiased radio surveys for UCDs remains stubbornly low, at $\sim$7-10\%, independent of instrumentation and survey volume.

On the other hand, \citet{kao16} recently conducted a survey of late L and T type UCDs, selecting brown dwarfs that exhibit previously known activity indicators, such as H$\alpha$ emission or periodic near-infrared variability.  This survey enjoyed a success rate of 80\% for six targets, appearing to be the most promising observing strategy to date.  The authors have attributed their success to an improved understanding of the emission phenomenology: that both radio and H$\alpha$ are emitted in the auroral regions of largely neutral atmospheres \citep{hal15}, not in stellar chromospheric-like or coronal-like structures \citep{ber01,ber05,ber10,bur05,ost06,lan07,lyn15}.  However, the foundation of this more efficient radio detection method is still an inefficient, and costly H$\alpha$ survey that depends on unbiased, targeted near-infrared/optical variability surveys to determine the absence or presence of magnetic activity indicators such as H$\alpha$.  Such near-infrared/optical surveys enjoy a success rate of $\sim$9\% for L4 to T8 dwarfs \citep{pin16}, that is statistically equivalent to the radio survey success rate for later-type UCDs.  This suggests that regardless of the theory of how UCDs flare, any temporally costly survey, whether using radio or near-infrared/optical spectroscopy, is required in order to generate new detections of magnetically active, later-type brown dwarfs.

By now, enough UCDs have been surveyed that the detection statistics for flaring radio emission can be compared in detail to those accumulated by \citet{pin16} in their search for H$\alpha$-emitting brown dwarfs.  They found that this H$\alpha$ detection rate as a function of spectral type is 67/195 ($\sim$34\%) for L0 to L9 dwarfs and 3/42 ($\sim$7.1\%) for T0 to T8 dwarfs.  Among these L dwarfs, H$\alpha$ emission is much more likely for types L0 to L3, where 60/120 ($\sim$50\%) are active, while only 7/75 ($\sim$9.3\%) of L4 to L9 dwarfs have detected red optical emission.  This trend may be indicative of chromospheric-like emission that ceases to function near L4/L5 due to lower temperatures (T$_{eff}\sim$ 1400 K) and increasingly neutral atmospheres \citep{kir05,rod15,pin16}.  Meanwhile, the fraction of each type that appears to be radio flaring is 3/36 of M dwarfs ($\sim$8\%), 3/61 of L dwarfs ($\sim$5\%), and 4/39 of T dwarfs ($\sim$10\%), suggesting a constant detection probability across spectral types, although obviously this apparent trend is based on a very small number of objects.  We note that our compiled statistics describe UCDs that would be detected from a survey sensitive to the rapidly varying radio emission detectable with the Arecibo radio telescope, where objects that were not detected in a previous radio survey may be revisited in future surveys.  Of course, instrumentation that is sensitive to quiescent emission, as can be found at VLA, has resulted in nearly double this detection rate, but at the cost of being insensitive to rapid temporal changes in the flares.  Although the overall detection probability for both H$\alpha$ and radio surveys are the same for L4 to T8 spectral types, H$\alpha$ surveys are more advantageous for M7 to L3 spectral types.

Interestingly, although the red optical integration times used during the \citet{pin16} survey were 900 - 1800 s, while the flaring radio emission sources were observed for $\sim$2 hr, the detection rates are similar for later spectral types.  Since the H$\alpha$ emission has a longer duty cycle than the flaring radio emission, this suggests that the H$\alpha$ emission regions have a larger total filling factor, indicating that these magnetic activity signatures are probing different magnetic phenomenon.  Moreover, the large discrepancy between the H$\alpha$ and radio-flaring detection rates for M7 to L3 dwarfs reinforces this notion.

Surveys that re-examine previously observed, but undetected, objects are important due to the varying levels of magnetic activity in these objects, a fact suggested by the non-detection of radio emission from J1047+21 using VLA \citep{ber06}, but subsequent detection of flaring radio emission from this same object using Arecibo \citep{rou12}.  Similarly, whereas no H$\alpha$ emission was detected in 2MASS J00361617+1821104 \citep{ber05}, later re-observation found clear H$\alpha$ emission \citep{pin16}.

\subsection{Rotation and Radio Luminosity Trends in UCD Magnetism}

\citet{rou16} reported a fundamental rotational period for J1122+25 of 0.288 hr, although the second and third subharmonics of this period also fit the data.  As the brown dwarf could be rotating close to its break-up velocity, and the mass of the object is constrained such that M$\leq$80 M$_J$, the radius must be R$\leq$0.9 R$_J$, and the rotational velocity is likely v$>$ 125 km s$^{-1}$.  With a radio flare luminosity of $>5.1 \times 10^{24}$ ergs s$^{-1}$, J1122+25 is more energetic than many flaring UCDs, which may be related to its rapid rotation.  The connection between rotation and activity was explored by \citet{mcl12} through the analysis of the results of their VLA radio survey of 104 M and L dwarfs, augmented by results from the literature.  Their search for a trend connecting radio luminosity and rotational velocity among M and L dwarfs was inconclusive, although it included only a handful of ultracool dwarfs, with only three known radio-emitting L dwarfs and no known radio-loud T dwarfs.  They did note, however, that there appeared to be few slowly rotating UCDs ($v~sin~i <$ 30 km s$^{-1}$) with smaller ($\nu$L$_{\nu}\lesssim$10$^{23}$ ergs s$^{-1}$) radio luminosities.  Furthermore, while studying the evolution of radio luminosity fraction as a function of Rossby number (${\rm Ro}$ = P/$\tau_{c}$) among stars of spectral types G0 to L3.5, where $P$ is the rotation period and $\tau_{c}$ is the convective overturn time, they showed that smaller Rossby numbers are correlated with increased radio luminosity.

Unfortunately, even by pooling the recent L and T brown dwarf radio detections (\citet{rou13} (and references therein), \citet{kao16}, \citet{lyn16}, this work), the relationship between rotational velocity and radio luminosity among UCDs still appears inconclusive, although suggestive that flaring radio luminosity fraction ($\nu$L$_{\nu}$/L$_{bol}$) may be weakly correlated with rotational velocity. Obviously, the ultra-rapid rotation of J1122+25, with its very small Rossby number and relatively large radio luminosity, would follow the general trend described in \citet{mcl12}. The ability to determine an empirical trend for UCDs alone, though, depends critically on the confirmation of the rapid rotation of J1122+25, its subsequent measurement, and a larger population of radio-emitting UCDs to examine.

Through examination of Figures 3 and 4, we can speculate about the radio luminosity evolution of brown dwarfs with spectral type, although our analysis admittedly suffers from too few known radio emitters for every brown dwarf spectral type.  A local maximum in $\nu$L$_{\nu}$/L$_{bol}$ may exist for early type L dwarfs near L0 - L2.  However, this peak in radio flaring luminosity corresponds to the binary source J0746+20 AB, and therefore represents two closely orbiting sources with a tangled and complicated magnetic field topology, as opposed to the apparently single sources that comprise the majority of the UCD population plotted.  This suggests that future UCD binaries may be untangled from single sources of the same spectral class on account of their elevated activity levels, as is also the case for J1314+13 AB.

Another potential trend in the data is the apparent decline in radio luminosity in both Figures 3 and 4, stretching from L2 to T3.  Since the trend also appears in Figure 3, where L$_{bol}$ is removed from the ordinate axis, this suggests that the apparent decline does not merely reflect the cooler temperatures and lower bolometric luminosities present in later spectral types.  We note, however that this possible trend relies on few data points and in many cases, only a single radio measurement per spectral type.

Interestingly, beyond T3, both quiescent and flaring radio luminosities appear to rise, which would confound expectations of a smooth, monotonic decline in radio activity between late M dwarf stars and Jupiter.  Similarly, if the decline in $\nu$L$_{\nu}$ from L2 to T3 is merely an artifact of too few detected radio sources, then UCD radio luminosity still appears to be relatively constant from M7 to T6.5, again in contrast to expectations.  Obviously, either trend is speculative since only a pair of late T dwarfs have been detected to date.  \citet{aud07} first noted that radio luminosity is approximately constant for UCDs, thereby indicating that a similar radio emission mechanism must operate in all radio-loud UCDs.  Accordingly, the emitting source does not weaken at cooler T$_{eff}$, and the radio emitting structures may have similar sizes.

If true, such trends would pose challenges for current models of brown dwarf interiors.  For example, based on convection-driven, geodynamo model scaling arguments, \citet{chr09} hypothesized that a steady decline in magnetic field strength, and presumably, magnetic activity, occurs on account of the reduced mass and internal heat, and thus, internal energy, that is available to power a dynamo.  This would indicate that a smooth transition in magnetic behavior occurs from rapidly rotating stars, through gas giant planets, to the terrestrial magnetized planets. Yet \citet{kao16} have also challenged the predictive power of this empirical ``law'' as the magnetic energy in J1047+21 appears to be significantly larger than predicted.  Finally, Figures 3 and 4 indicate that future detections of UCDs beyond T7 should show markedly reduced flaring and quiescent radio emission, with steeply declining magnetic field strengths for substellar objects between J1047+21, and radio-luminous gas giant planets such as Jupiter.

\section{Conclusion}
The second survey conducted at a center frequency of 4.75 GHz at Arecibo Observatory has examined 27 UCDs in total in a search for new sources of periodic radio flaring.  Twenty of these objects were previously unobserved at Arecibo Observatory, and 19 of these unobserved by any research group.  We have detected radio emission from J1047+21, J1122+25, and J1314+13 AB, and failed to detect emission from J1439+19.  Although we tentatively reported J1439+19 as a potential radio-flaring object in \citet{rou13}, our subsequent detection of more flares associated with this object suggest RFI that has been aliased with the terrestrial sidereal rotation rate.  While J1047+21 and J1314+13 AB have been previously observed to exhibit flaring behavior, our detections of flares from these two sources reveals new properties of their radio emission.  The detection of a new radio-emitting T dwarf, J1122+25, as initially reported in \citet{rou16} is the most rapidly rotating UCD detected to date, with a computed period of 0.288 hr (or the second or third subharmonics of this period).  This source is also the most radio luminous T dwarf detected thus far, a fact that could be attributed to the influence of its ultra-fast rotation on its internal magnetic dynamo.

Further study of J1122+25, in particular, is required on account of its apparent ultra-rapid rotation.  A search for its quiescent radio emission, as well as other magnetic activity indicators, such as H$\alpha$ emission as predicted by \citet{kao16}, would serve to evaluate the hypothesized correlation between these two indicators.  A measurement of its projected rotational velocity would improve our understanding of how radio luminosity evolves with rotational velocity (and thus, Rossby number) for ultracool dwarfs (i.e. \citet{mcl12}).  If possible, long term monitoring of J1122+25's radio flaring behavior could determine whether the brown dwarf experiences differential rotation (i.e. \citet{wol14}), which may or may not be suppressed due to the presence of strong magnetic fields as alleged by theoretical models \citep{bro08}.  Both parameters, the rotational velocity and the differential rotation rate, if any, of J1122+25 would have profound implications for the development and evaluation of dynamo models.

Future survey work should attempt to populate the region between the L8 flaring source 2MASS J10430758+2225236 \citep{kao16} and the recent T dwarf detections.  It would be helpful to contextualize these detections to learn if they represent unique UCDs that are uncommon in some way, such as being extremely rapid rotators, or if they are typical and representative of the late T UCD population.  In addition, T dwarfs later than T6.5, Y dwarfs, and hot, young exoplanets with brown dwarf-like luminosities (e.g. \citet{bon13}, \citet{opp13}) are tempting targets to search for radio emission that will allow us to unlock the secrets of their magnetic fields, and thereby probe their interiors.  These observations would permit the examination of how magnetic strengths and structures evolve from the well-studied M7 to L5 UCDs, to the less studied L5 to T6.5 brown dwarfs, to the entirely unstudied T6.5 to Y0 region, and beyond.  Further discoveries in these regimes would provide valuable information and constraints to guide the development of giant exoplanetary dynamo models \citep{san04}, and evaluate efforts to understand stellar dynamo theory, through the generalization of Solar dynamo models to rapidly rotating, and fully convective stars (e.g. \citet{brown08}).

\section{Acknowledgements}

MR acknowledges support from the Center for Exoplanets and Habitable Worlds and the NASA Pennsylvania Space Grant Consortium Fellowship. The Center for Exoplanets and Habitable Worlds is supported by the Pennsylvania State University, and the Eberly College of Science. The Arecibo Observatory is operated by SRI International under a cooperative agreement with the National Science Foundation (AST-1100968), and in alliance with Ana G. M\'{e}ndez-Universidad Metropolitana, and the Universities Space Research Association.  This research has made use of NASA's Astrophysics Data System and the SIMBAD database, operated at CDS, Strasbourg, France.

\clearpage

\begin{deluxetable}{lcccccc}
\tabletypesize{\scriptsize}
\tablecolumns{7}
\tablecaption{Survey Target Properties}
\tablehead{
	\colhead{Name}&
	\colhead{Spectral Type}&
	\colhead{Distance}&
	\colhead{v $sin i$}&
	\colhead{Properties}&
	\colhead{Radio Flux}&
	\colhead{Radio}\\
	\colhead{}&
	\colhead{}&
	\colhead{(pc)}&
	\colhead{(km s$^{-1}$)}&
	\colhead{References}&
	\colhead{Density ($\mu$Jy)}&
	\colhead{References}
}
\startdata
2MASS J07003664+3157266	& L3.5$\phantom{p}$ & 12.2 & 29.9 & 1,2 & $<$78,$<$1455 & 3,4\\
2MASS J07271824+1710012 & T8$\phantom{.0p}$ & $\phantom{0}$9.1 & & 5,6 & $<$54,$<$1101 & 3,4\\
SDSS J074149.15+235127.5 & T5$\phantom{.0p}$ & 23.6 & & 7,8 & & \\
SDSS J074201.41+205521.5 & T5$\phantom{.0p}$ & 15.0 & & 7,9 & & \\
WISEPA J075004+272545 & T8.5$\phantom{p}$ & 15.8 & & 6,10 & & \\
SDSS J075547.87+221215.6 & T5$\phantom{.0p}$ & 18.0 & & 5 & & \\
2MASS J08105865+1420390 & M9$\phantom{.0p}$ & 20.3 & & 11 & $<$39 & 12\\
2MASS J0825196+211552 & L7.5$\phantom{p}$ & 10.7 & & 13,14 & $<$45,$<$1242 & 4,15\\
WISE J083811.45+151115.1 & T6.5$\phantom{p}$ & 20.0 & & 16 & & \\
SDSS J090023.68+253934.3 & L7$\phantom{.0p}$ & 24.8 & & 17,18 & & \\
SDSS J092308.70+234013.7 & L1$\phantom{.0p}$ & 21.4 & & 18 & & \\
2MASS J09373487+2931409 & T6$\phantom{.0p}$ & $\phantom{0}$6.1 & & 5,6 & $<$66,$<$1227 & 3,4\\
SDSS J104307.51+222523.5 & L8$\phantom{.0p}$ & 17.2 & & 19 & & \\
SDSS J104335.08+121314.1 & L7$\phantom{.0p}$ & 14.6 & & 20,9 & & \\
2MASS J10475385+212434 & T6.5$\phantom{p}$ & 10.6 & & 21,5,22 & 2700 & 23\\
SDSS J110401.29+195922.3 & L4$\phantom{.0p}$ & 18.8 & & 24 & & \\
WISE J111838.70+312537.9 & T8.5$\phantom{p}$ & $\phantom{0}$8.3 & & 25 & & \\
WISEPC J112254.73+255021.5 & T6$\phantom{.0p}$ & 16.9 & & 5 & & \\
WISEPC J121756.91+162640.2 & T9$\phantom{.0p}$ & $\phantom{0}$6.7 & & 5 & & \\
SDSS J121951.45+312849.4 & L8$\phantom{.0p}$ & 18.1 & & 20,18 & & \\
Ross 458C\tablenotemark{a} & T8.5p & 11.7 & & 26,27,28 & & \\
2MASS J13004255+1912354 & L1$\phantom{.0p}$ & 13.9 & & 11,29 & $<$87 & 15\\
ULAS 130217.21+130851.2 & T8.5$\phantom{p}$ & 20.1 & & 30,10 & & \\
2MASS J13142039+1320011 AB & M7$\phantom{.0p}$ & 17.2 & 45 & 31,32 & 8000 & 33 \\
PSO J201.0320+19.1072\tablenotemark{b} & T3.5$\phantom{p}$ & 20.0 & & 34 & & \\
ULAS J133553.45+113005.2 & T9$\phantom{.0p}$ & 10.3 & & 35,36 & $<$1242 & 4\\
2MASS J1439284+192915 & L1$\phantom{.0p}$ & 14.4 & 11.1 & 37,14 & 1062 (RFI) & 4\\
\enddata
\tablecomments{{\bf References.} (1) \citet{tho03}; (2) \citet{bla10}; (3) \citet{ant13}; (4) \citet{rou13}; (5) \citet{bur02}; (6) \citet{kir11}; (7) \citet{kna04}; (8) \citet{loo07}; (9) \citet{fah12}; (10) \citet{kir12}; (11) \citet{giz00}); (12) \citet{pha07}; (13) \citet{kir00}; (14) \citet{dah02}; (15) \citet{ber06}; (16) \citet{abe11}; (17) \citet{zha09}; (18) \citet{sch10}; (19) \citet{cru07}; (20) \citet{chi06}; (21) \citet{bur99}; (22) \citet{vrb04}; (23) \citet{rou12}; (24) \citet{cru03}; (25) \citet{wri13}; (26) \citet{gol10}; (27) \citet{scho10}; (28) \citet{bur11}; (29) \citet{sch07}; (30) \citet{bur10}; (31) \citet{mcl11}; (32) \citet{for16}; (33) \citet{wil15}; (34) \citet{dea11}; (35) \citet{bur08}; (36) \citet{mar10}; (37) \citet{kir99}.  This table benefited from ``The M, L, T, and Y dwarf compendium,'' DwarfArchives.org, 6 Dec. 2002 and from the ''List of Brown Dwarfs,'' johnstonsarchive.net, 2012.}
\tablenotetext{a}{Also known as ULAS 130042+122115}
\tablenotetext{b}{Also known as 2MASS J13240776+1906271}
\end{deluxetable}
\clearpage

\begin{deluxetable}{lcccccc}
\tabletypesize{\scriptsize}
\tablecolumns{7}
\tablecaption{Survey Detection Results}
\tablehead{
	\colhead{Object}&
	\colhead{Spectral Type}&
	\colhead{Time on Source}&
	\colhead{Detected Flux}&
	\colhead{$\nu$L$_{\nu}$}&
	\colhead{L$_{bol}$\,\tablenotemark{a}}&
	\colhead{$\nu$L$_{\nu}$/L$_{bol}$}\\
	\colhead{}&
	\colhead{}&
	\colhead{(ks)}&
	\colhead{Density (mJy)}&
	\colhead{(log L$_{\odot}$)}&
	\colhead{(log L$_{\odot}$)}&
	\colhead{(log L$_{\odot}$)}
}
\startdata
2MASS J07003664+3157266 & L3.5$\phantom{p}$ & $\phantom{0}$8.4 & $<$1.546 & $<$-8.991 & $\phantom{<}$-3.96$\phantom{{*}}$ & $<$-5.031\\
2MASS J07271824+1710012 & T8$\phantom{.0p}$ & $\phantom{0}$9.0 & $<$1.564 & $<$-9.242 & $\phantom{<}$-5.26$\phantom{{*}}$ & $<$-3.982\\
SDSS J074149.15+235127.5 & T5$\phantom{.0p}$ & $\phantom{0}$7.8 & $<$1.443 & $<$-8.447 & $\phantom{<}$-4.82{*} & $<$-3.627\\
SDSS J074201.41+205520.5 & T5$\phantom{.0p}$ & 11.4 & $<$1.419 & $<$-8.846 & $\phantom{<}$-4.82{*} & $<$-4.026\\
WISEPA J0753003.84+272544.8 & T8.5$\phantom{p}$ & $\phantom{0}$7.8 & $<$1.029 & $<$-8.943 & $<$-5.58{*} & $<$-3.363\\
SDSS J075547.87+221215.6 & T6$\phantom{.0p}$ & 10.2 & $<$1.389 & $<$-8.699 & $\phantom{<}$-5.01{*} & $<$-3.689\\
2MASS J08105865+1420390 & M9$\phantom{.0p}$ & $\phantom{0}$9.0 & $<$1.311 & $<$-8.620 & $\phantom{<}$-3.39{*} & $<$-5.230\\
SDSS J082519.45+211550.3 & L7.5$\phantom{p}$ & $\phantom{0}$7.2 & $<$1.262 & $<$-9.193 & $\phantom{<}$-5.21{*} & $<$-3.983\\
WISE J083811.45+151115.1 & T6.5$\phantom{p}$ & $\phantom{0}$6.6 & $<$1.973 & $<$-8.455 & $\phantom{<}$-5.13{*} & $<$-3.325\\
SDSS J090023.68+253934.3 & L6$\phantom{.0p}$ & $\phantom{0}$6.6 & $<$1.906 & $<$-8.283 & $\phantom{<}$-4.34{*} & $<$-3.943\\
SDSS J092308.70+234013.7 & L1$\phantom{.0p}$ & $\phantom{0}$6.6 & $<$4.785 & $<$-8.012 & $\phantom{<}$-3.67{*} & $<$-4.342\\
2MASS J09373487+2931409 & T7$\phantom{.0p}$ & $\phantom{0}$6.6 & $<$2.111 & $<$-9.454 & $\phantom{<}$-5.26{*} & $<$-4.194\\
SDSS J104307.51+222523.5 & L8$\phantom{.0p}$ & $\phantom{0}$7.2 & $<$2.652 & $<$-8.458 & $\phantom{<}$-4.55{*} & $<$-3.908\\
SDSS J104335.08+121314.1 & L7$\phantom{.0p}$ & $\phantom{0}$7.2 & $<$1.118 & $<$-8.975 & $\phantom{<}$-4.44{*} & $<$-4.535\\
2MASS J10475385+2124234 & T6.5$\phantom{p}$ & 84.6 & $\phantom{<}$1.880 & $\phantom{<}$-9.508 & $\phantom{<}$-5.35$\phantom{{*}}$ & $\phantom{<}$-4.158\\
SDSS J110401.29+195922.3 & L4$\phantom{.0p}$ & 12.6 &$<$1.381 & $<$-8.664 & $\phantom{<}$-4.09{*} & $<$-4.574\\
WISE J111838.70+312537.9 & T8.5$\phantom{p}$ & $\phantom{0}$7.2 & $<$1.129 & $<$-9.463 & $<$-5.58{*} & $<$-3.883\\
WISEPC J112254.73+255021.5 & T6$\phantom{.0p}$ & 68.4 & $\phantom{<}$2.860 & $\phantom{<}$-8.878 & $\phantom{<}$-5.01{*} & $\phantom{<}$-3.868\\
WISEPC J121756.91+162640.2 & T9$\phantom{.0p}$ & $\phantom{0}$7.2 & $<$2.371 & $<$-9.325 & $<$-5.58{*} & $<$-3.745\\
SDSS J121951.45+312849.4 & L8$\phantom{.0p}$ & $\phantom{0}$7.2 & $<$1.028 & $<$-8.825 & $\phantom{<}$-4.55{*} & $<$-4.275\\
Ross 458C\tablenotemark{b} & T8.5$\phantom{p}$ & 10.2 & $<$1.394 & $<$-9.072 & $\phantom{<}$-5.61{*} & $<$-3.462\\
2MASS J13004255+1912354 & L1$\phantom{.0p}$ & $\phantom{0}$7.2 & $<$2.312 & $<$-8.702 & $\phantom{<}$-4.12{*} & $<$-4.582\\
ULAS J130217.21+130851.2 & T8.5$\phantom{p}$ & 10.2 & $<$1.445 & $<$-8.586 & $<$-5.58{*} & $<$-3.006\\
2MASS J13142039+1320011 AB & M7$\phantom{.0p}$ & $\phantom{0}$7.2 & $\phantom{<}$1.080 & $\phantom{<}$-8.848 & $\phantom{<}$-3.17{*} & $\phantom{<}$-5.678\\
PSO J201.0320+19.1072\tablenotemark{c} & T3.5$\phantom{p}$ & $\phantom{0}$7.2 & $<$1.105 & $<$-8.707 & $\phantom{<}$-4.62{*} & $<$-4.087\\
ULAS J133553.45+113005.2 & T9$\phantom{.0p}$ & $\phantom{0}$7.2 & $<$1.420 & $<$-9.171 & $<$-5.58{*} & $<$-3.591\\
2MASS J1439284+192915 & L1$\phantom{.0p}$ & 45.0 & $<$1.398 & $<$-8.892 & $\phantom{<}$-3.67{*} & $<$-5.222\\
\enddata
\tablecomments{Although J1047+21 was detected previously \citep{rou12}, this table lists the flux density of the newly detected flare. The potential radio emission of J1439+19 reported in Table 2 \citep{rou13} should be amended with the upper limit provided here.}
\tablenotetext{a}{Asterisks denote bolometric luminosities inferred from \citet{vrb04}. Bolometric luminosities for objects later than T8 are not computed in \citet{vrb04}; thus, the minimum given bolometric luminosity (-5.58, for T8) is used instead.}
\tablenotetext{b}{Also known as ULAS 130042+122115}
\tablenotetext{c}{Also known as 2MASS J13240776+1906271}
\end{deluxetable}
\clearpage

\begin{figure}
\centering
\includegraphics[width=1.1\textwidth,angle=0]{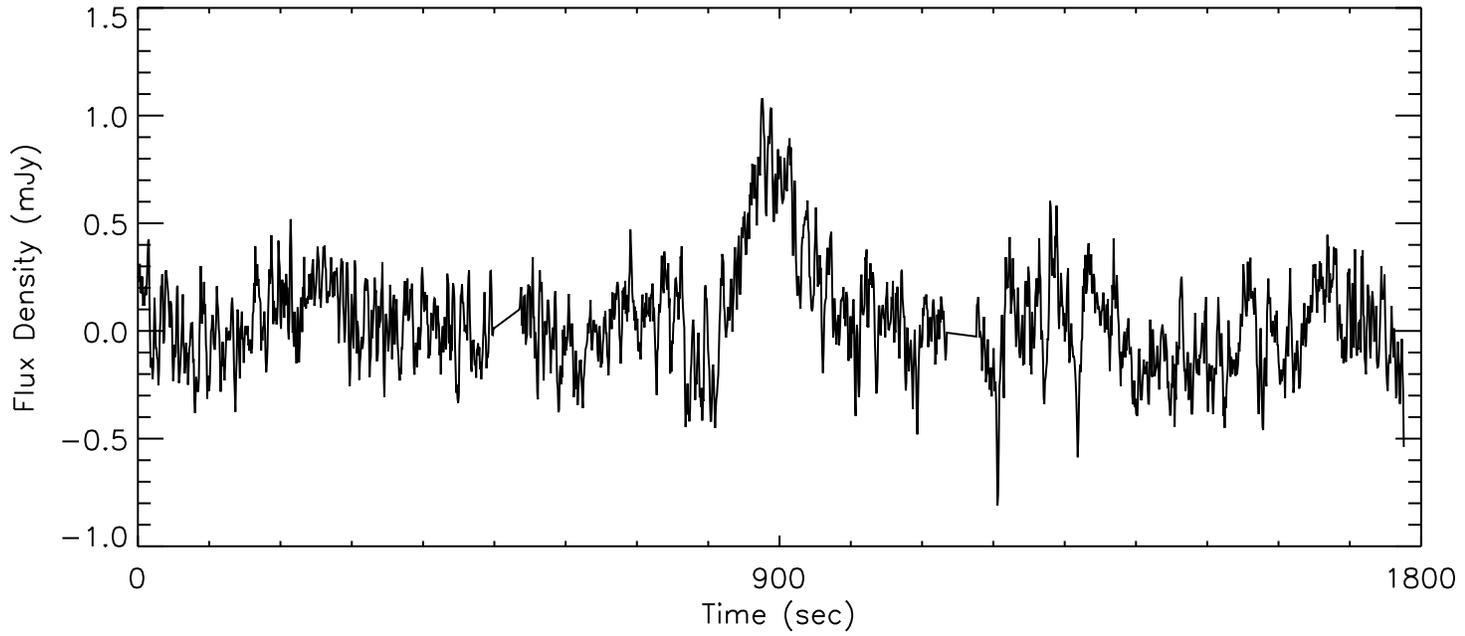}
\caption{The temporal Stokes V profile of a J1314+13 AB flare, averaged over the entire $\sim$1 GHz bandpass, as recorded on 2013 May 4.  The 57\% left circularly polarized flare reaches a peak flux density of $\sim$1 mJy. The diagonal line segments near 500 s and 1200 s show where calibration caused gaps in the data collection. The profile has been binned to 0.9 s resolution, with 3-point smoothing. \label{fig2}}
\end{figure}

\begin{figure}
\centering
\includegraphics[width=1.1\textwidth,angle=-90,scale=0.7]{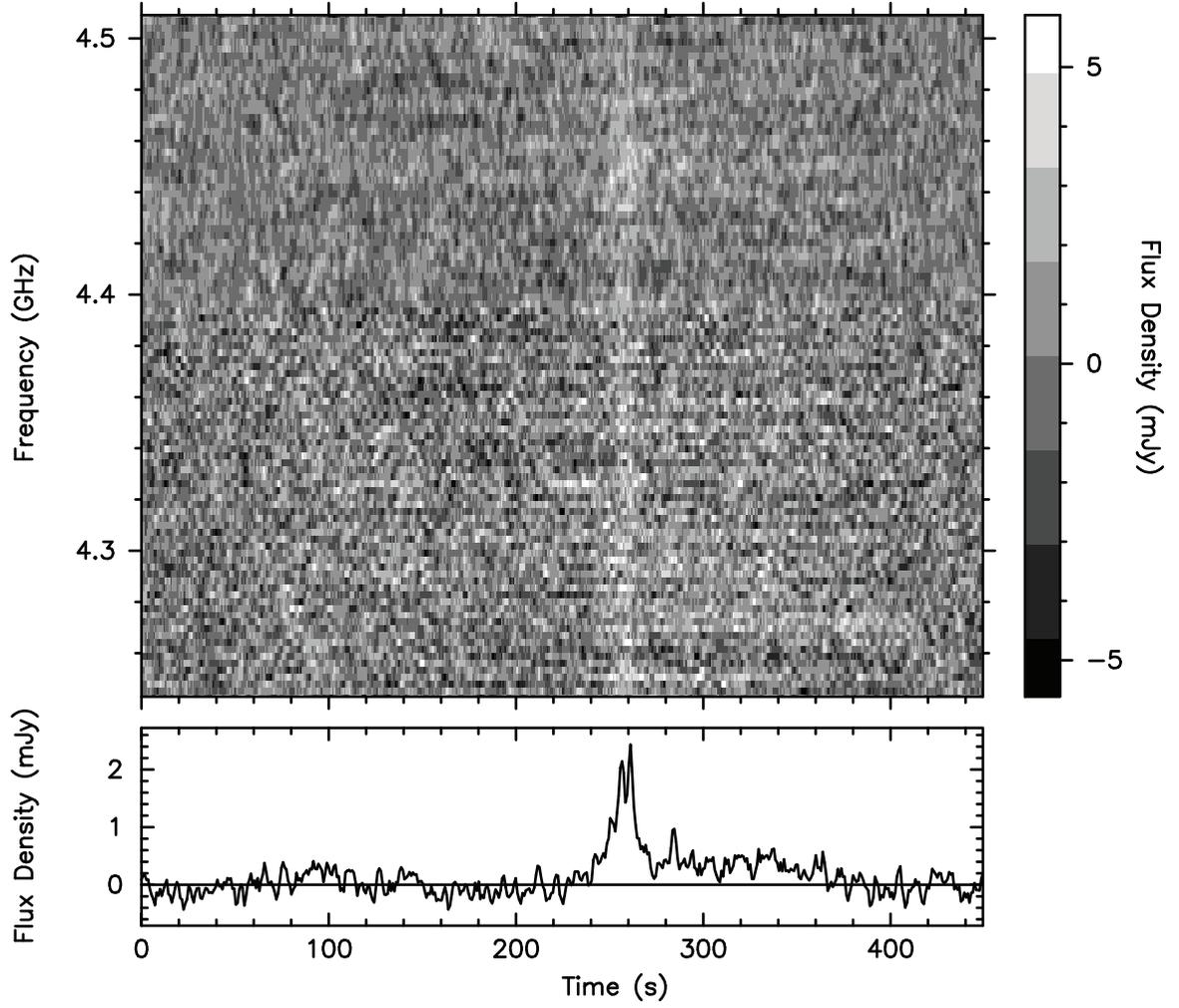}
\caption{The dynamic spectrum (top) and average temporal profile (bottom) of the Stokes V radio burst from J1047+21 detected on 2013 May 6.  The dynamic spectra has been smoothed to a 2 MHz frequency resolution and binned to 6 s temporal resolution, to enhance the flare's visibility.  The increased graininess in the dynamic spectrum below 4.4 GHz is due to extra, low-level noise generated by the data acquisition hardware.  The temporal profile has been integrated over the $\sim$500 MHz spectrometer bandpass and binned to 0.9 s resolution.\label{fig1}}
\end{figure}

\begin{figure}
\centering
\includegraphics[width=1.1\textwidth,angle=0]{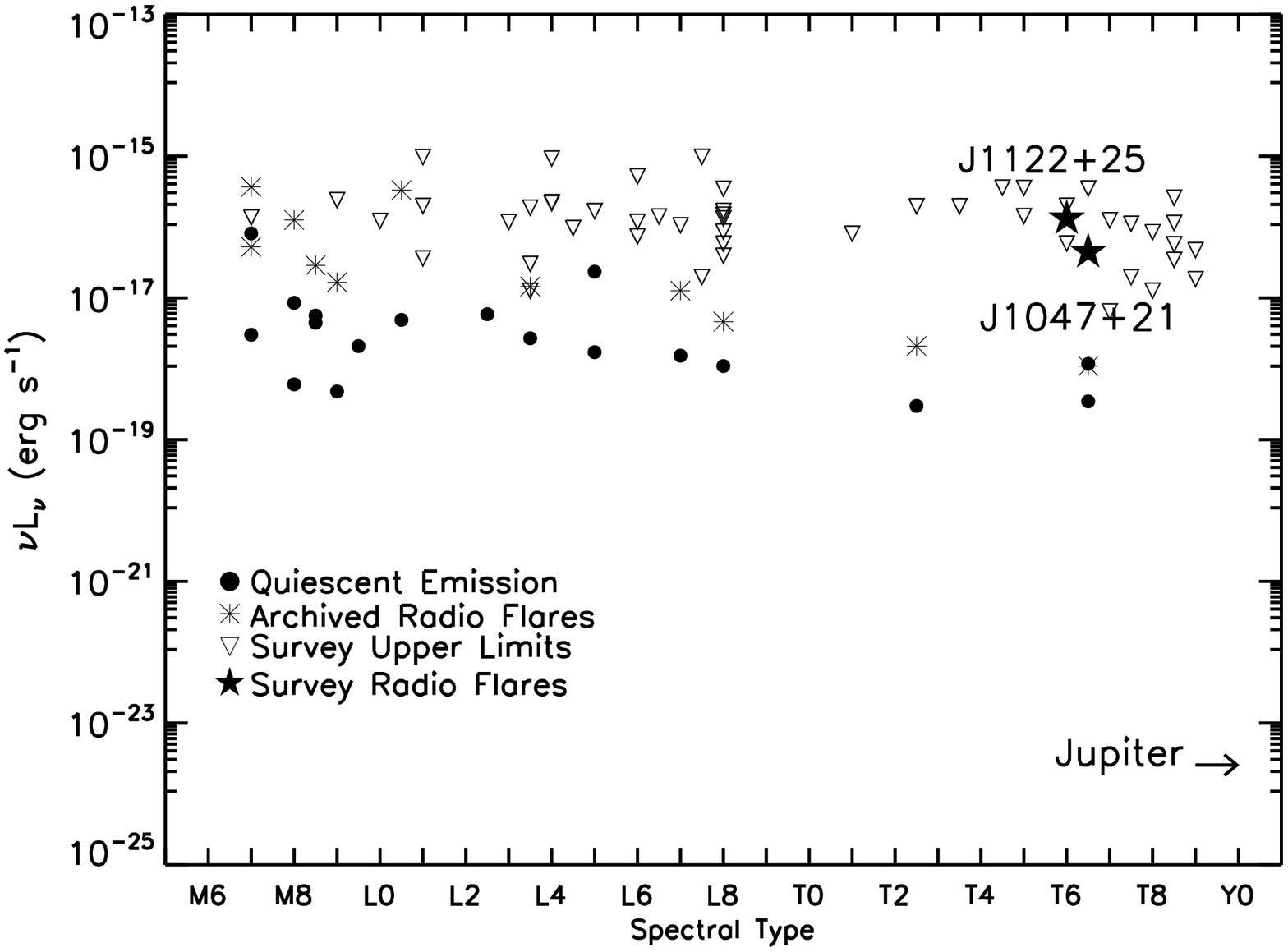}
\caption{Logarithmic plot of the radio luminosity versus spectral type of detected radio-emitting UCDs \citep{mcl12,bur13,kao16,lyn16,rou13,rou16}. The recent detections of J1122+25\citep{rou16} and J1047+21\citep{rou12} are denoted with filled stars, while inverted open triangles represent the upper limits for the survey targets from both the current Arecibo survey (Table 2) and our previous survey \citep{rou13}.  For non-detected sources observed in both surveys, we plot the most sensitive upper limit.  Note that Jupiter's radio emission \citep{gui05,laz07} places it off the diagram to the lower right, potentially indicating that a large range of substellar radio luminosities remain unexplored.  \label{fig3}}
\end{figure}

\begin{figure}
\centering
\includegraphics[width=1.1\textwidth,angle=0]{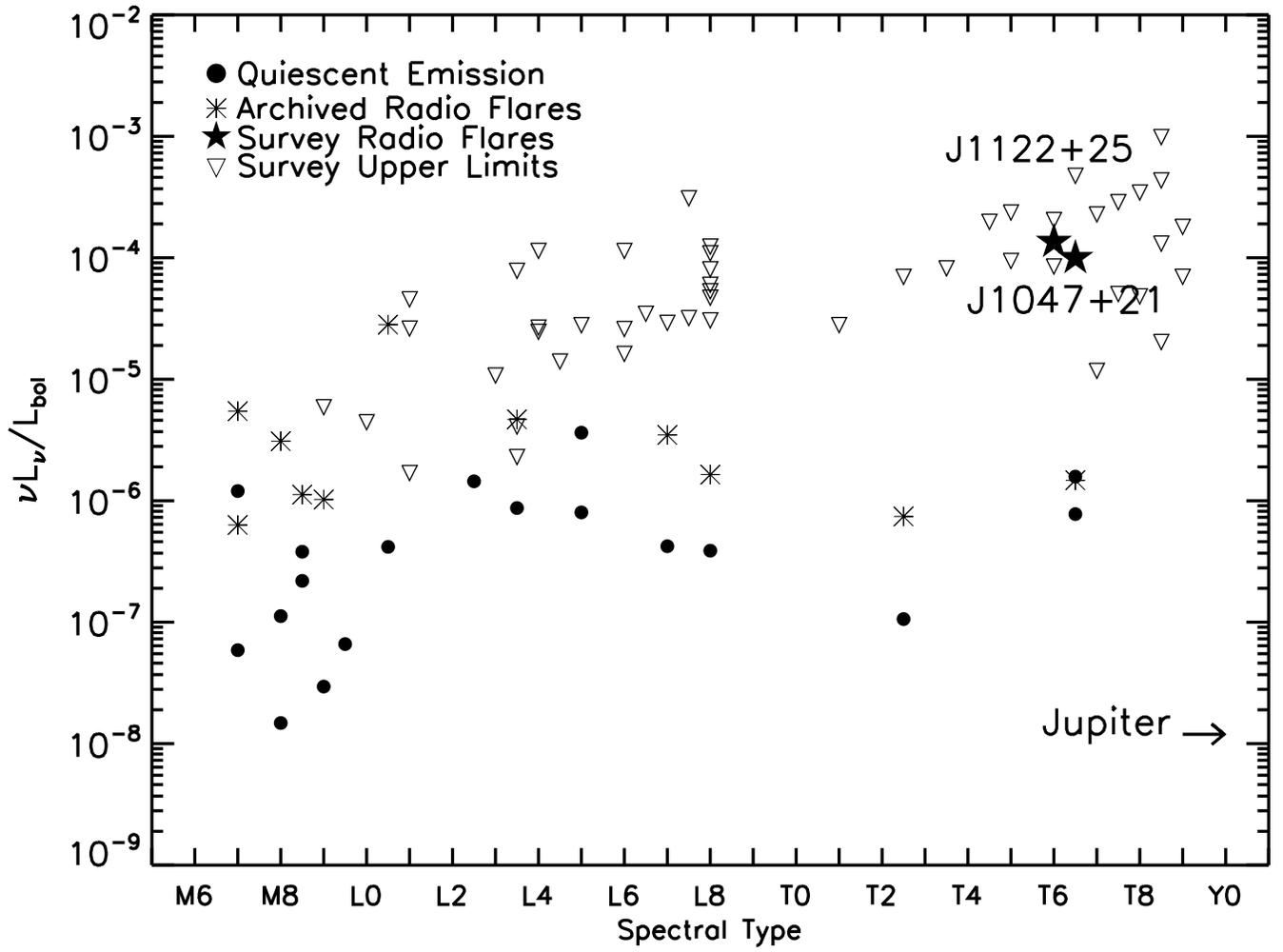}
\caption{Similar to Figure 3, we examine the evolution of the radio luminosity fraction of UCDs, by considering the ratio of their quiescent and flaring radio luminosities to their bolometric luminosities, as a function of spectral type.  A general trend of declining radio activity with spectral type is apparent after L0 to L2, but is interrupted by the activity of the two Arecibo-detected late T dwarfs, J1122+25 and J1047+21. \label{fig4}}
\end{figure}

\end{document}